\documentclass[prb,reprint,superscriptaddress]{revtex4-2}

\usepackage{amsmath}
\usepackage{bm}
\usepackage{graphicx}
\usepackage{xcolor}
\usepackage{hyperref}

\begin{document}

\title{Photon echo from Maki-Thompson interference in a fluctuating s-wave superconductor}

\date{\today}

\author{Hui Zhou}
\affiliation{Institute of Physics, Chinese Academy of Sciences, Beijing 100190, China}
\affiliation{Songshan Lake Materials Laboratory, Dongguan, Guangdong 523808, China}

\author{Yuan Wan}
\email{yuan.wan@iphy.ac.cn}
\affiliation{Institute of Physics, Chinese Academy of Sciences, Beijing 100190, China}
\affiliation{Songshan Lake Materials Laboratory, Dongguan, Guangdong 523808, China}

\begin{abstract}
We study theoretically the nonlinear optical response of a fluctuating s-wave superconductor driven by two successive terahertz pulses. We show that the response exhibits a photon echo signal that arises from the quantum interference between the time-reversed electron and hole trajectories in the Maki-Thompson process. By contrast, the transport of the fluctuating Cooper pairs responsible for the Aslamazov-Larkin conductivity does not generate an echo. The photon echo signal can be readily measured using terahertz two-dimensional coherent spectroscopy, highlighting its potential to distinguish different fluctuation conductivity mechanisms.
\end{abstract}

\maketitle

\section{Introduction}

\begin{figure}
\includegraphics[width = \columnwidth]{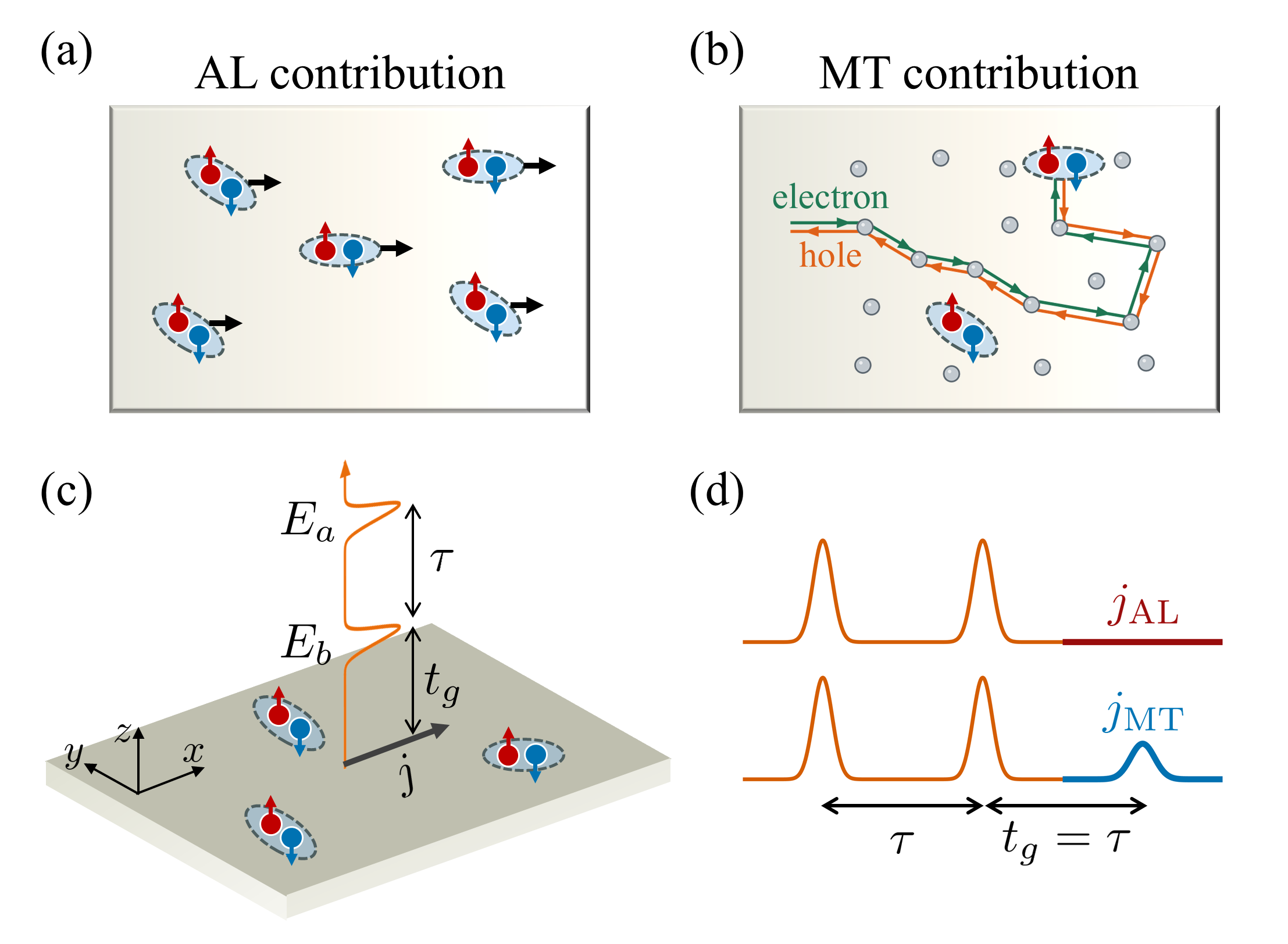}
\caption{(a) The transport of the fluctuating Cooper pairs in an electric field gives rise to the Aslamazov–Larkin (AL) contribution to the fluctuation conductivity. (b) The Maki-Thompson (MT) contribution to the fluctuation conductivity arises from the quantum interference between the electron and hole trajectories related by time reversal. (c) The set up used in this work, where an s-wave superconductor thin film above its transition temperature is driven by two terahertz pulses separately by a time $\tau$. We monitor the optical response encoded in the time-dependent current $\bm{j}$ as a function of the gating time $t_g$ after the second pulse. (d) Our analysis reveals that the MT process produces a photon echo, whereas the AL process does not.}
\label{fig:sketch}
\end{figure}

Fluctuating superconductivity is at the heart of several prominent problems in condensed-matter physics~\cite{Larkin2005}. In the cuprates, superconducting correlations above the transition temperature are closely connected to the pseudogap and anomalous transport~\cite{Emery1995,Timusk1999}, while in disordered superconducting films, enhanced fluctuations govern the approach to the superconductor–insulator transition and the nearby metallic regime~\cite{Goldman1998,Sacepe2011,Kapitulnik2019}. More broadly, in low dimensional and disordered systems, pairing correlations may develop well before long-range superconducting order is established~\cite{Emery1995,Sacepe2011}. Understanding the electrodynamic response of this fluctuating regime consequently remains an active line of research.

A prominent manifestation of superconducting fluctuations is the excess conductivity above the transition temperature $T_c$, known as the fluctuation conductivity. In an s-wave superconductor, this response contains contributions from several microscopic processes~\cite{Larkin2005}. Above $T_c$, the fluctuating Cooper pairs may accelerate under the influence of the electric field, thereby producing an excess conductivity known as the Aslamazov–Larkin (AL) contribution~\cite{Aslamazov1968} (Fig.~\ref{fig:sketch}a). More subtle is the Maki-Thompson (MT) contribution~\cite{Maki1968,Thompson1970}, where the excess conductivity stems from Andreev-like reflection of unpaired electrons from superconducting fluctuations (Fig.~\ref{fig:sketch}b). 

Although the AL and MT contributions originate from distinct microscopic processes, disentangling them experimentally remains challenging. The established approach relies on magnetoconductivity measurements~\cite{Clarke1971,LopesdosSantos1985,Bergmann1984,Gordon1985}. It exploits the distinct magnetic-field dependences of the two contributions to separate them through fits to transport data. While this method has provided crucial insights into fluctuation conductivity, it requires a quantitative comparison with a theoretical model. A complementary probe based on their intrinsic dynamical properties would therefore be valuable.

In this work, we show that the photon echo provides a diagnostic that can \emph{qualitatively} distinguish the two fluctuation conductivity channels. A photon echo is a nonlinear optical signal generated by two phase coherent pulses~\cite{Kurnit1964,Abella1966}. Following the arrival of the second pulse, the signal emerges after a delay equal to the inter-pulse delay. As the inter-pulse delay increases, it appears correspondingly later, analogous to an acoustic echo. The photon echo is intimately related to quantum interference during the system’s dynamical evolution~\cite{Mukamel1995,Hamm2011}. Therefore, the presence or absence of an echo can reveal quantum interference structures, or the lack thereof, hidden in the linear response~\cite{Wan2019,Li2021,LiWan2024}.

This observation suggests a natural way to distinguish the fluctuation conductivity channels. The MT contribution arises from quantum interference between a pair of electron and hole trajectories related by time reversal. A two-pulse drive may generate a photon echo by modulating this quantum interference. On the other hand, the AL contribution arises from the transport of fluctuating Cooper pairs and does not require interference. Consequently, the AL contribution should not exhibit an echo.

Our theoretical analysis confirms these expectations. We calculate the optical response encoded in the electric current induced by two successive terahertz pulses (Fig.~\ref{fig:sketch}c). We denote the delay time between the pulses by $\tau$, and refer to the time of measurement following the second pulse as the gating time $t_g$. The MT contribution exhibits a photon echo that appears at $t_g = \tau$,  whereas the AL contribution does not (Fig.~\ref{fig:sketch}d). The echo signal is a manifestation of the quantum interference inherent to the MT process: When $t_g \neq \tau$, the interference is suppressed by the pulses, whereas it is reinstated when $t_g = \tau$. Furthermore, the echo fades away as the pulse delay time $\tau$ increases, mirroring the dephasing process of the electron. Thus, the photon echo is a direct, time-domain probe of electron phase coherence in fluctuating superconductors. Importantly, the predicted photon echo signal can be measured using recently developed terahertz two-dimensional coherent spectroscopy~\cite{Woerner2013,Lu2018,Chen2025,Huang2026} in a standard experimental configuration, providing a practical route to its detection.

This mechanism parallels the photon echo response found in a previous study of weak localization by one of us~\cite{Li2024}. There, the echo originated from the quantum interference between a self-intersecting electron trajectory and its time-reversed partner, providing a time-domain interferometric probe of weak localization. The present work extends this trajectory-based interpretation of photon echo to a fluctuating superconductor. The relevant time-reversed objects are now electron and hole trajectories embedded in the MT process. The analogy suggests that photon echoes can serve more broadly as probes of quantum interference contributions to transport.

The past two decades have seen a surge of interest in the nonlinear optical response of superconductors. We now position our work in this broader body of research. Most previous studies have considered the superconducting state, with particular emphasis on the Higgs mode~\cite{Pekker2015,Shimano2020,Matsunaga2013,Matsunaga2014,Tsuji2015,Cea2016,Huang2025}. Above $T_c$, a series of theoretical works has investigated the steady state nonlinear optical response~\cite{Kovalev2021,Parafilo2022,Sonowal2023,Miranda2026}. Ref.~\onlinecite{Lemonik2018} investigates the nonequilibrium dynamics of the superconducting fluctuations and their manifestation in the transient optical response following a quench. Ref.~\onlinecite{Fumarola2007} considers an echo in the evolution following a local density perturbation in a paired Fermi gas above $T_c$; this phenomenon is distinct from a photon echo.

On the experimental front, femtosecond pump-probe studies identified a narrow fluctuation regime above $T_c$ in cuprates~\cite{Demsar1999}, linear terahertz spectroscopy characterized the corresponding fluctuation time scale~\cite{Bilbro2011}, and three-pulse measurements distinguished pairing-related and phase-coherence dynamics~\cite{Madan2014}. Ultrafast mid-infrared measurements in cuprates~\cite{Perfetti2015} and optical-pump/terahertz-probe measurements in NbN~\cite{Antonio2026} further revealed a pronounced slowing of the dynamics on approaching the transition. It would be interesting to put our predictions to test on a standard sample such as aluminum film.

The rest of this work is organized as follows. In Sec.~\ref{sec:model}, we set up the model used in our theoretical analysis. In Sec.~\ref{sec:results}, we show through explicit calculations that the MT contribution exhibits a photon echo whereas the AL contribution does not. In Sec.~\ref{sec:interference}, we demonstrate that the photon echo arises from the quantum interference in the MT process. In Sec.~\ref{sec:discussion}, we discuss several open questions. 

\section{Model \label{sec:model}}

We consider a disordered s-wave superconductor film above its superconducting transition temperature. We work in the regime where all relevant time scales, including the electron phase coherence time $\tau_\phi$ and the Ginzburg-Landau time $\tau_{GL}$, are much longer than the electron elastic scattering time $\tau_{el}$. Equivalently, the electron phase-coherence length $l_\phi$ and the superconducting correlation length $\xi_{GL}$ are much larger than the electron mean free path $l$. In this \emph{diffusive} regime, the fluctuating superconductor is characterized by a handful of constants whereas microscopic details such as the band structure drop out from the description.

We assume the film thickness $d$ is smaller than or comparable to the superconducting correlation length $\xi_{GL}$ as well as the electron phase-coherence length $l_\phi$. Therefore, the system can be viewed as effectively two-dimensional. Furthermore, we take the film width to be infinite. Extending our analysis beyond the two-dimensional limit is straightforward. We shall briefly discuss this aspect in Sec.~\ref{sec:discussion}.

A terahertz electromagnetic wave is incident on the film. Its polarization is parallel to the film. The electromagnetic wave's impact on the film is encoded in a time-dependent, spatially uniform electric field $\bm{E}(t)$. In what follows, we use the Coulomb gauge: $\bm{E}(t) = -\partial\bm{A}/\partial t$, where $\bm{A}(t)$ is the vector potential. The scalar potential $\phi = 0$.

The dynamics of the superconducting order parameter $\Delta$ above the transition temperature are modeled by the linearized stochastic Ginzburg-Landau equation~\cite{Larkin2005} (we use natural units $k_B = \hbar = 1$ unless stated otherwise):
\begin{align}
\partial_t \Delta = [D(\bm{\nabla}-2ie_0\bm{A})^2 - \tau^{-1}_{GL}]\Delta + \eta.
\label{eq:TDGL}
\end{align}
Here, $D>0$ is the electron diffusion constant in the normal state. $e_0=-|e_0|$ is the electron charge. $\bm{A}(t)$ is the aforementioned vector potential~\footnote{Strictly speaking, the stochastic Ginzburg-Landau equation should be expressed in the so-called $\mathcal{K}$-gauge~\cite{Kamenev1999,Feigelman2000,Levchenko2007}. In the present context, the scalar potential $\phi = 0$, and the charge accumulation does not occur. The $\mathcal{K}$-gauge coincides with the Coulomb gauge.}. The Ginzburg-Landau time $\tau_{GL}$ depends on the temperature through the universal relation:
\begin{align}
\tau_{GL} = \frac{\pi}{8T\ln(T/T_c)} \approx \frac{\pi}{8}\frac{1}{T-T_c}.
\label{eq:tau_GL}
\end{align}
Here, $T_c$ should be understood as the mean-field transition temperature. The second expression is valid when $T$ is sufficiently close to $T_c$. $\eta$ is the complex Gaussian white noise field. Its correlation function is:
\begin{align}
\langle \eta(x,t)\eta^\ast (x,t') \rangle = 2T\delta^{(2)}(x-x')\delta(t-t').
\end{align}
The linearized Ginzburg-Landau equation captures only the Gaussian fluctuations of the superconducting order parameter. The linearization is no longer valid when non-Gaussian fluctuations become significant.

The central quantity of interest is the time-dependent, spatially uniform sheet current density $\bm{j}(t)$ induced by the electromagnetic wave. The current radiates an electromagnetic signal $\bm{E}_{sig}(t) = Z_0 \bm{j}(t)/2$ that can be measured by time-domain terahertz spectroscopy. $Z_0 \approx 377$ Ohm is the free-space wave impedance.

$\bm{j}(t)$ consists of three contributions~\cite{Levchenko2007}, 
\begin{align}
\bm{j}(t) = \bm{j}_D (t) + \bm{j}_{AL} (t) + \bm{j}_{MT} (t).
\label{eq:j_total}
\end{align} 
$\bm{j}_D$ is the Drude response in the normal state.  Over time scales much longer than the elastic scattering time $\tau_{el}$, the Drude response is instantaneous:
\begin{align}
\bm{j}_D(t) = \sigma_D \bm{E}(t).
\end{align}
$\sigma_D>0$ is the Drude conductivity. 

$\bm{j}_{AL}$ is the contribution from the fluctuating supercurrent:
\begin{align}
\bm{j}_{AL}(t) = 4e_0D\mathrm{Im} \langle \Delta^\ast (\bm{\nabla}-2ie\bm{A}) \Delta\rangle.
\label{eq:j_AL}
\end{align}
Eq.~\eqref{eq:j_AL} may be inferred from the Ginzburg-Landau equation Eq.~\eqref{eq:TDGL} by drawing an analogy with single-particle quantum mechanics. $\langle\cdots\rangle$ denotes the statistical average. In the linear response regime, the supercurrent contribution yields the celebrated AL conductivity, and hence the subscript.

Finally, $\bm{j}_{MT}$ reflects the correction to the normal state Drude conductivity by the MT process, where the unpaired electrons experience Andreev-like reflection by the fluctuating superconducting order parameter:
\begin{align}
\bm{j}_{MT}(t) = \int^t_{-\infty} K(t,t')\bm{E}(t')dt'.
\label{eq:j_MT}
\end{align}
The kernel $K$ is local in space but nonlocal in time:
\begin{align}
K(\bm{x},t,t') = 4e^2_0D\int d^2\bm{y} d^2\bm{z}\, \mathcal{C}^{\frac{t+t'}{2}}_{t-t',0}(\bm{x}-\bm{y}) 
\nonumber\\
 \times \langle \Delta(\bm{y},\frac{t+t'}{2})\Delta^\ast(\bm{z},\frac{t+t'}{2})\rangle \overline{\mathcal{C}}^{\frac{t+t'}{2}}_{0,t-t'}(\bm{z}-\bm{x}) .
 \label{eq:K_MT}
\end{align}
Spatial translational symmetry implies that $K$ is in fact independent of $\bm{x}$. $\mathcal{C}$ and $\overline{\mathcal{C}}$ are the retarded and advanced Cooperons. In the context of the MT process, $\mathcal{C}$ and $\overline{\mathcal{C}}$ describe the quantum interference of the time-reversed trajectories of an electron and a hole. They obey the Cooperon equation~\cite{Altshuler1982}:
\begin{subequations}
\begin{align}
[2\partial_t - D(\bm{\nabla}-ie_0\bm{A}^s_t)^2 + \tau^{-1}_\phi] \mathcal{C}^s_{tt'}(\bm{x}-\bm{x}')
\nonumber\\
 = \delta^{(2)}(\bm{x}-\bm{x}') \delta(t-t'),
\end{align}
and
\begin{align}
[-2\partial_t - D(\bm{\nabla}-ie_0\bm{A}^s_t)^2 + \tau^{-1}_\phi]\overline{\mathcal{C}}^s_{tt'}(\bm{x}-\bm{x}')
\nonumber\\
 = \delta^{(2)}(\bm{x}-\bm{x}') \delta(t-t').
\end{align}
\label{eq:cooperon_equation}
\end{subequations}
The shorthand notation $\bm{A}^s_t = \bm{A}(s+t/2) + \bm{A}(s-t/2)$. $\tau_\phi$ is a phenomenological phase coherence time for the electron. In Sec.~\ref{sec:interference}, we provide a heuristic justification for Eq.~\eqref{eq:K_MT}. 

Eq.~\eqref{eq:TDGL}, Eq.~\eqref{eq:j_total}, and Eq.~\eqref{eq:cooperon_equation} form a closed set of equations that govern the electromagnetic response of the fluctuating superconductor driven by a uniform, time-dependent vector potential. Formally, they are obtained by expanding the nonlinear sigma model~\cite{Feigelman2000} for the disordered s-wave superconductor to quadratic order in the Cooperon and diffuson channels and then integrating out those modes. We refer the interested reader to Ref.~\onlinecite{Levchenko2007} for the derivation. 

We close this section by commenting on the applicability of the above system of equations. First, as we have mentioned above, the linearized Ginzburg-Landau equation Eq.~\eqref{eq:TDGL} captures only the Gaussian fluctuations of the superconducting order parameter. Second, we have omitted the so-called density-of-states correction to the conductivity~\cite{Ami1978,Altshuler1983}. Akin to the Drude response, it is instantaneous in time, and, therefore, does not yield any interesting nonlinear response for our purpose. It is also sub-leading to the AL and MT contribution in the regime we are interested in. Third, we have also dropped the weak localization contribution to the current density since it has been analyzed in the previous work~\cite{Li2024}. We shall discuss it briefly in Sec.~\ref{sec:discussion}.

\section{Results \label{sec:results}}

In this section, we present the main results of this work. In Sec.~\ref{sec:solving}, we derive the nonlinear AL and MT response to an arbitrary time-dependent electric field. In Sec.~\ref{sec:echo}, we specialize to the case of two terahertz pulses and demonstrate the presence and absence of a photon echo in the MT and AL contribution, respectively. We also apply our theory to the aluminum film to make contact with experiments. 

\subsection{Solving the equations \label{sec:solving}}

We calculate the sheet current density $\bm{j}(t)$ induced by an arbitrary time-dependent electric field $\bm{E}(t)$. Since the AL contribution and the MT contribution add, we treat them separately. We omit the Drude contribution at the outset as it does not produce any nonlinear response.

We first consider the AL contribution. Since the vector potential is spatially uniform, the Ginzburg-Landau equation Eq.~\eqref{eq:TDGL} is diagonal in momentum space:
\begin{align}
\partial_t \Delta(\bm{q}) =  -[D(\bm{q}-2e_0\bm{A}(t))^2 + \tau^{-1}_{GL}] \Delta(\bm{q}) + \eta(\bm{q}),
\end{align}
where $\langle \eta(\bm{q},t) \eta^\ast (\bm{q}',t')\rangle = 2T\delta(t-t')\delta_{\bm{q}\bm{q}'}$. We define the second moments,
\begin{align}
L(\bm{q},t) = \langle \Delta^\ast(\bm{q},t)\Delta(\bm{q},t)\rangle.
\end{align}
It is easy to show that $L(\bm{q},t)$ obeys the equation:
\begin{align}
\partial_t L(\bm{q}) = -2[D(\bm{q}-2e_0\bm{A}(t))^2 + \tau^{-1}_{GL}]L(\bm{q}) + 2T.
\end{align}
with the initial condition $L(\bm{q},t\to -\infty) = 0$. Solving the above differential equation, we obtain:
\begin{align}
L(\bm{q},t) = 2T\int^t_{-\infty} e^{- 2\int^t_{s}[D(\bm{q}-2e_0\bm{A}(u))^2+\frac{1}{\tau_{GL}}]du} ds.
\label{eq:L_q}
\end{align}

In the momentum space, the AL contribution reads:
\begin{align}
\bm{j}_{AL}(t) = 4e_0D\int \frac{d^2\bm{q}}{(2\pi)^2}(\bm{q}-2e_0\bm{A}(t)) L(\bm{q},t).
\end{align}
Substituting Eq.~\eqref{eq:L_q} and carrying out the integration over $\bm{q}$, we find:
\begin{align}
\bm{j}_{AL}(t) = \frac{2e^2_0T}{\pi}\int^t_{-\infty} \frac{e^{-\frac{2(t-s)}{\tau_{GL}}-f(t,s)}}{t-s}
\nonumber\\
\times (\overline{\bm{A}}(t,s) - \bm{A}(t))ds.
\label{eq:j_AL_result}
\end{align}
Here, $\overline{\bm{A}}(t,s)$ is the moving average of the vector potential over the interval from $s$ to $t$:
\begin{align}
\overline{\bm{A}}(t,s) = \frac{1}{t-s}\int^t_s \bm{A}(u) du.
\end{align}
$f(t,s)$ is the coherence factor for the AL contribution:
\begin{align}
f(t,s) = 8e^2_0D\int^t_s (\bm{A}(u) - \overline{\bm{A}}(t,s))^2 du.
\end{align}
$f(t,s)>0$ encodes the suppression of the AL current by a dynamical vector potential. 

We then turn to the MT contribution. Thanks to the spatial translational symmetry, the Cooperon equations are diagonal in momentum space as well:
\begin{subequations}
\begin{align}
[2\partial_t + D(\bm{q} - e_0\bm{A}^s_{t})^2 + \tau^{-1}_\phi] \mathcal{C}^s_{tt'}(\bm{q}) = \delta(t-t');
\\
[-2\partial_t + D(\bm{q} - e_0\bm{A}^s_{t})^2 + \tau^{-1}_\phi] \overline{\mathcal{C}}^s_{tt'}(\bm{q}) =  \delta(t-t').
\end{align}
\end{subequations}
The solutions are:
\begin{subequations}
\begin{align}
\mathcal{C}^s_{tt'}(\bm{q}) = \theta(t-t')e^{-\frac{1}{2}\int^t_{t'} [D(\bm{q}-e_0\bm{A}^s_{u})^2+\frac{1}{\tau_\phi}] du};
\\
\overline{\mathcal{C}}^s_{tt'}(\bm{q}) = \theta(t'-t)e^{-\frac{1}{2}\int^{t'}_{t} [D(\bm{q}-e_0\bm{A}^s_{u})^2+\frac{1}{\tau_\phi}] du}.
\end{align}
\label{eq:cooperon_q}
\end{subequations}
$\theta(\cdot)$ denotes the Heaviside step function. 

In the momentum space, the kernel for the MT contribution reads:
\begin{align}
K(t,t') = 2e^2_0D\int \frac{d^2\bm{q}}{(2\pi)^2} \mathcal{C}^{\frac{t+t'}{2}}_{t-t',0}(\bm{q}) L(\bm{q},\frac{t+t'}{2}) 
\nonumber\\
\times \overline{\mathcal{C}}^{\frac{t+t'}{2}}_{0,t-t'}(\bm{q}).
\end{align}
Substituting Eq.~\eqref{eq:cooperon_q} and Eq.~\eqref{eq:L_q}, and integrating over $\bm{q}$, we obtain:
\begin{align}
K(t,t') = \frac{e^2_0T}{\pi}e^{-\frac{t-t'}{\tau_\phi}}\int^{\frac{t+t'}{2}}_{-\infty} \frac{e^{-\frac{t+t'-2s}{\tau_{GL}}-g(t,t',s)}}{t-s} ds.
\label{eq:K_result}
\end{align}
Here, we have defined a coherence factor for the MT process:
\begin{align}
g(t,t',s) = 8e^2_0D\int^t_s [\bm{W}(u) - \overline{\bm{W}}(t,s)]^2 du.
\end{align}
$\bm{W}$ is an effective vector potential defined with respect to fixed $t$ and $t'$:
\begin{align}
\bm{W}(u) = \left\{\begin{array}{cc}
\frac{1}{2}[\bm{A}(u)+\bm{A}(t+t'-u)] & (u>\frac{t+t'}{2}) \\
\bm{A}(u) & (u<\frac{t+t'}{2})
\end{array}\right. .
\label{eq:W_def}
\end{align}
$\overline{\bm{W}}$ is its moving average:
\begin{align}
\overline{\bm{W}}(t,s) = \frac{1}{t-s}\int^t_s \bm{W}(u) du.
\end{align}

Eq.~\eqref{eq:j_AL_result} and Eq.~\eqref{eq:K_result} are the main results of this section. It is interesting to compare their forms. For both, the nonlinear responses are encoded in their respective coherence factors $f(t,t')$ and $g(t,t',s)$. Crucially, the vector potential enters $g(t,t',s)$ through a time-reversed combination $\bm{A}(u) + \bm{A}(t+t'-u)$. This structure is our first hint for the photon echo in the MT contribution.

As a sanity check of Eq.~\eqref{eq:j_AL_result} and Eq.~\eqref{eq:K_result}, we compute the DC conductivity due to the AL and MT contribution. To this end, we set $\bm{E}(t) = \bm{E}$. It follows that $\bm{A}(t) = -\bm{E}t$. For linear response, we may drop the coherence factors $f$ and $g$. The AL contribution thus reads:
\begin{align}
\bm{j}_{AL}(t) = \frac{e^2_0T}{\pi}\bm{E}\int^t_{-\infty} e^{-\frac{2(t-s)}{\tau_{GL}}}ds = \frac{e^2_0T}{2\pi}\tau_{GL}\bm{E},
\end{align}
which yields the AL conductivity: 
\begin{align}
\sigma_{AL} = \frac{e^2_0T}{2\pi}\tau_{GL} \approx \frac{e^2_0}{16}\frac{T_c}{T-T_c}.
\end{align}
In the second equality, we have inserted the expression for the Ginzburg-Landau time $\tau_{GL}$ near $T_c$. The kernel for the MT contribution reads:
\begin{align}
K(t-t') = \frac{e^2_0T}{\pi} e^{-\frac{t-t'}{\tau_\phi}}\int^{\frac{t+t'}{2}}_{-\infty} \frac{e^{-\frac{t+t'-2s}{\tau_{GL}}}}{t-s} ds 
\nonumber\\
= \frac{e^2_0T}{\pi} e^{-\frac{t-t'}{\tau_\phi}}e^{\frac{t-t'}{\tau_{GL}}}E_1(\frac{t-t'}{\tau_{GL}}).
\end{align}
$E_1(x) = \int^\infty_x e^{-y}/y\, dy$ is the exponential integral. Integrating the kernel $K$ over the time difference $t-t'$ yields the MT conductivity:
\begin{align}
\sigma_{MT}  = \int^\infty_0 K(t) dt = \frac{e^2_0T}{\pi}\frac{\tau_\phi\tau_{GL}}{\tau_\phi-\tau_{GL}}\ln(\frac{\tau_\phi}{\tau_{GL}})
\nonumber\\
 \approx \frac{e^2_0}{8}\frac{T_c}{T-T_c}\ln(\frac{\tau_\phi}{\tau_{GL}}).
\end{align}
In the second line, we have assumed $\tau_\phi \gg \tau_{GL}$ and inserted the expression for $\tau_{GL}$. We see that $\sigma_{MT}$ can be parametrically larger than $\sigma_{AL}$ in this case.

\subsection{Photon echo \label{sec:echo}} 

\begin{figure}
\includegraphics[width = \columnwidth]{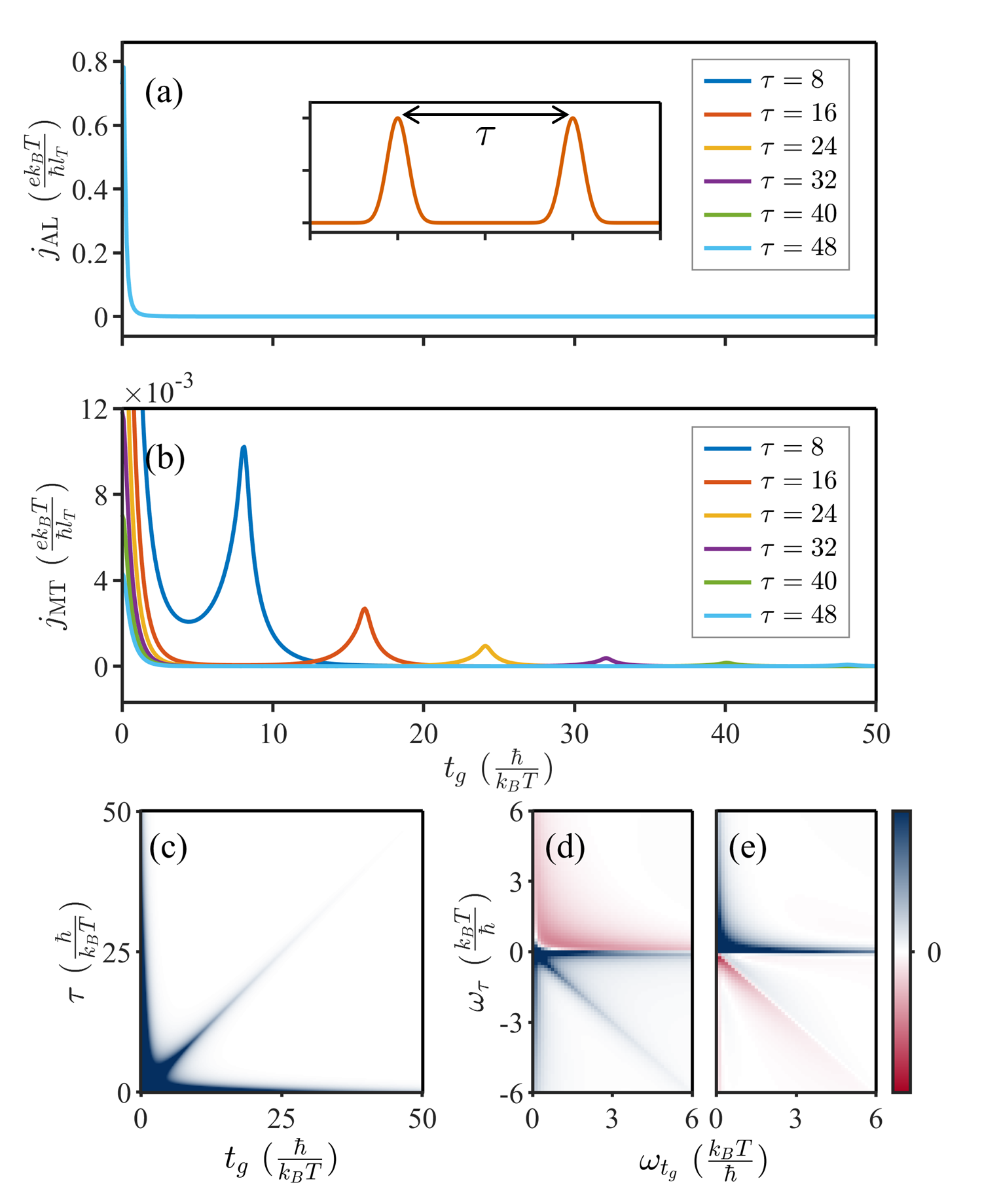}
\caption{Sheet current density from the AL contribution ($\bm{j}_{AL}$) and the MT contribution ($\bm{j}_{MT}$) in a fluctuating s-wave superconductor induced by two successive Gaussian electromagnetic pulses. See the main text for the parameter choices. (a) $\bm{j}_{AL}$ as a function of the gating time $t_g$ for various representative pulse delay times $\tau$. Inset shows the pulse profile as well as the definition of the pulse delay time $\tau$. (b) Similar to (a) but for $\bm{j}_{MT}$. We have used the MT response due to the second pulse as the baseline, and subtracted it off from the full, two-pulse response to highlight the latter's temporal structure. (c) $\bm{j}_{MT}$ as a function of both $t_g$ and $\tau$. (d) The real part of the two-dimensional coherent spectrum, obtained by Fourier transforming the time-domain data shown in (c). (e) Same as (d) but for the imaginary part. In (d)(e), only the first and fourth quadrants are shown as the spectrum in the other half plane is related by complex conjugation. The data are linearly scaled such that the maximum magnitude is $1$.}
\label{fig:echo}
\end{figure}

We are now ready to calculate the nonlinear current response induced by the two terahertz optical pulses by applying the results from the previous section. Recall the expressions for the AL contribution $\bm{j}_{AL}$ and the MT contribution $\bm{j}_{MT}$ only depend on a handful of parameters: the electron diffusion constant $D$, the temperature $T$, the Ginzburg-Landau time $\tau_{GL}$, and the phase coherence time $\tau_\phi$. For presentation purposes, it is convenient to combine them with $\hbar$ and $k_B$ to form a set of natural units for the time, the sheet current density, and the electric field. The natural time unit is the ``Planck" time $\hbar/(k_BT)$. Combining the Planck time with the diffusion constant $D$ produces a thermal diffusion length $l_T = \sqrt{\hbar D/{k_BT}}$. It follows that the natural electric field scale is $k_BT/(el_T)$, and the natural sheet current density scale is $ek_BT/(\hbar l_T)$.

For simplicity, we use a Gaussian pulse profile:
\begin{align}
\bm{E}(t) = \bm{E}_a e^{-\frac{t^2}{2t^2_p}} +  \bm{E}_b e^{-\frac{(t-\tau)^2}{2t^2_p}},
\end{align}
where $t_p>0$ controls the pulse duration. $\tau>0$ is the delay time between the two pulses. $\bm{E}_{a,b}$ are the peak electric field strength.

Fig.~\ref{fig:echo}(a) and (b) show, respectively, the AL contribution $\bm{j}_{AL}$ and the MT contribution $\bm{j}_{MT}$. For $\bm{j}_{MT}$, we use the MT response due to the second pulse as the baseline, and subtract it off from the full, two-pulse response to highlight the latter's temporal structure, i.e. $\bm{j}_{MT} = \bm{j}_{MT}(\bm{E}_a,\bm{E}_b)-\bm{j}_{MT}(\bm{E}_b)$. We set the Ginzburg-Landau time $\tau_{GL} = 5 \hbar/(k_BT)$ and the phase coherence time $\tau_{\phi} = 25 \hbar/(k_BT)$. For the pulses, we set the pulse duration $t_p = 0.1 \hbar/(k_BT)$ and the peak electric fields $\bm{E}_a = \bm{E}_b = 10/\sqrt{2\pi}\, k_BT/(el_T)$. 

We observe a clear echo signal in the MT contribution $\bm{j}_{MT}$. For fixed pulse delay $\tau$, $\bm{j}_{MT}$ exhibits a peak at $t_g = \tau$. The echo fades away with increasing pulse delay time $\tau$. For the thin film geometry considered here, this echo signal in $\bm{j}_{MT}$ translates to a photon echo through the relation $\bm{E}_{sig} \propto \bm{j}$. By contrast, the AL contribution $\bm{j}_{AL}$ exhibits no echo signal. Therefore, the photon echo provides a \emph{qualitative} diagnostic to distinguish these two fluctuation conductivity channels. 

Scanning both $t_g$ and $\tau$ provides a complete picture of the MT contribution (Fig.~\ref{fig:echo}(c)). The echo appears as the diagonal feature in the $\tau-t_g$ plane. Fourier transforming the time-domain data produces the two-dimensional spectrum (Figs.~\ref{fig:echo}(d)(e)). The echo manifests itself as a highly anisotropic peak in the fourth quadrant: the peak is elongated in the antidiagonal direction ($\omega_\tau = -\omega_{t_g}$) but narrow in the diagonal direction ($\omega_\tau = \omega_{t_g}$).

\begin{figure}
\includegraphics[width = \columnwidth]{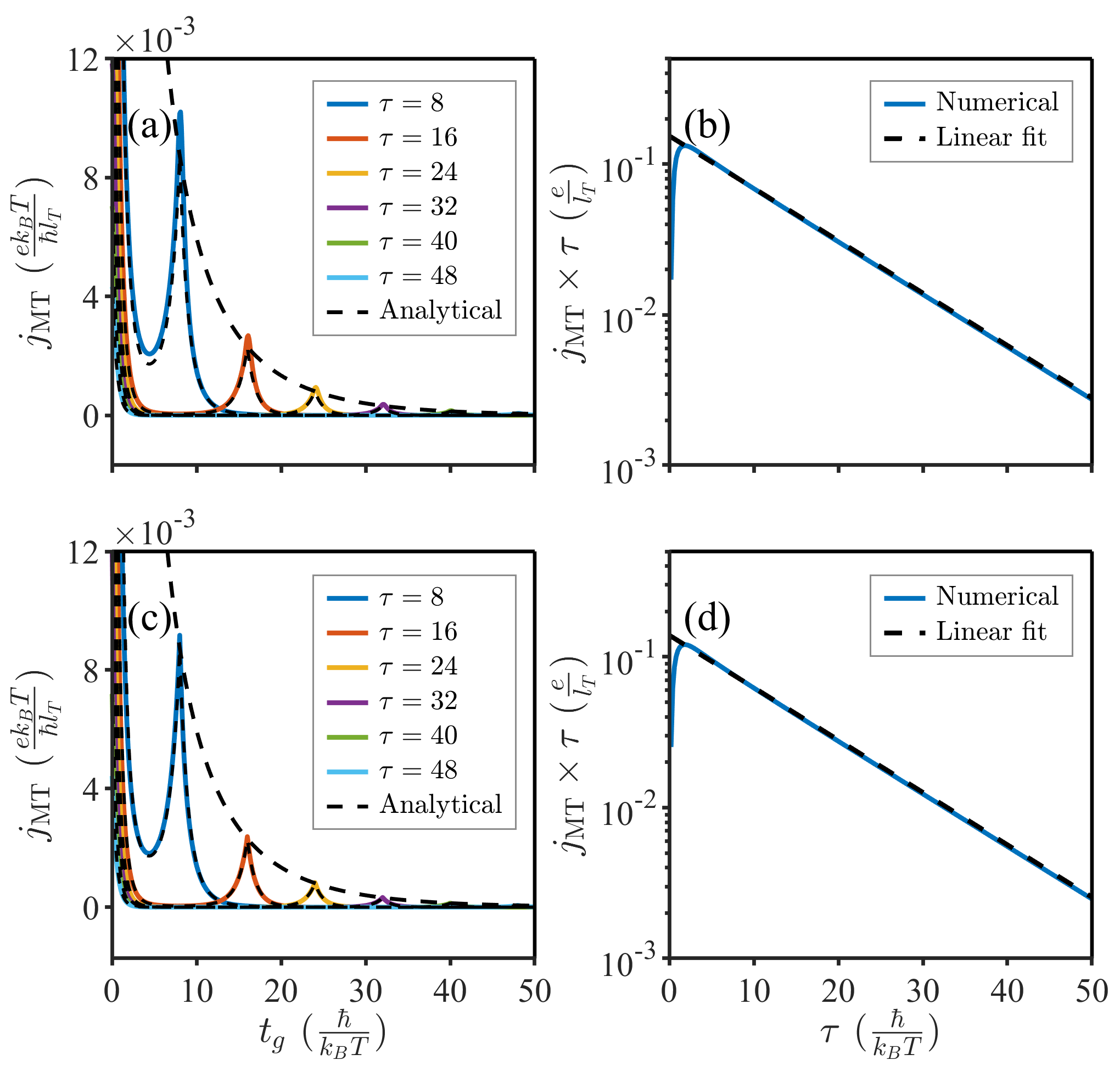}
\caption{Comparison of the numerically calculated MT contribution $\bm{j}_{MT}$ and the analytically obtained asymptotic form. (a) Solid lines show the numerically calculated $\bm{j}_{MT}$ as a function of the gating time $t_g$ for representative pulse delay times $\tau$, taken from Fig.~\ref{fig:echo}(b). The dashed lines show the analytical results. (b) The numerically calculated $j_{MT} \times \tau$ at the gating time $t_g=\tau$ as a function of $\tau$. The dashed line is the linear fit. (c)(d) Same as (a)(b) but for Gaussian pulses with reduced pulse duration $t_p =0.01 \hbar/(k_BT)$. Note that the linear fit in (d) has the same slope as that in (b).}
\label{fig:asymptotics}
\end{figure}

Having established the existence of photon echo in the MT contribution, we now analyze its asymptotic behavior in the limit of $\tau\to \infty$. In particular, we wish to clarify which factors control the decay of the echo signal. To this end, we assume the optical pulses are sufficiently short so that they can be approximated as Dirac-$\delta$ functions, $\bm{E}(t) \approx \sqrt{2\pi}\bm{E}_a t_p  \delta(t) +  \sqrt{2\pi}\bm{E}_b t_p  \delta(t-\tau)$. The factor $\sqrt{2\pi}\bm{E}_{a,b}t_p$ is the total area under the Gaussian pulse profile. A tedious but straightforward calculation shows:
\begin{align}
\bm{j}_{MT}(t_g) &=\frac{\sqrt{2\pi} e^2_0 T\bm{E}_at_p }{\pi(2\pi e^2_0 D\bm{E}_b^2t^2_p\frac{4\min\{t_g,\tau\}^2}{\tau+t_g}+\frac{\tau+t_g}{\tau_{GL}})}e^{-\frac{\tau+t_g}{\tau_\phi}}\nonumber\\
&\times\exp(-4\pi e^2_0D\bm{E}_b^2 t^2_p \frac{|\tau-t_g|}{\tau+t_g}\min\{t_g,\tau\}).
\label{eq:asymp}
\end{align}
Here, $\min\{t_g,\tau\}$ refers to the lesser of the two arguments. The above expression is valid in the limit of large $\tau$. The maximum of $\bm{j}_{MT}$, located at $t_g=\tau$, is given by: 
\begin{align}
\bm{j}_{MT}(\tau)=\frac{\sqrt{2\pi} e^2_0T\bm{E}_at_p}{\pi(2\pi e^2_0D\bm{E}_b^2 t^2_p+\frac{1}{\tau_{GL}})}\frac{e^{-\frac{2\tau}{\tau_\phi}}}{2\tau}.
\label{eq:asymp_echo}
\end{align} 
Eq.~\eqref{eq:asymp_echo} shows the echo directly probes the phase coherence. The coherence time $\tau_\phi$ can be extracted by measuring the decay of the echo amplitude as a function of the pulse delay.
 
Fig.~\ref{fig:asymptotics}(a) compares the numerical result, taken from Fig.~\ref{fig:echo}(b), against the asymptotic form Eq.~\eqref{eq:asymp}. We observe a small deviation of the numerically calculated $\bm{j}_{MT}$ from the analytically predicted behavior for moderate pulse delay time $\tau$. Nevertheless, the deviation diminishes as $\tau$ increases. The small discrepancy is expected because the asymptotic analysis uses Dirac delta pulses, whereas the numerical calculation uses Gaussian pulses with finite pulse duration $t_p$. Reducing the pulse duration down to $t_p = 0.01 \hbar/(k_BT)$ significantly improves the agreement between the two (Fig.~\ref{fig:asymptotics}(c)). Finally, regardless of the pulse duration, we find the magnitude of the echo signal, measured at $t_g = \tau$, exhibits $\exp(-2\tau/\tau_\phi)/\tau$ scaling as predicted (Figs.~\ref{fig:asymptotics}(b)(d)).

\begin{figure}
\includegraphics[width = \columnwidth]{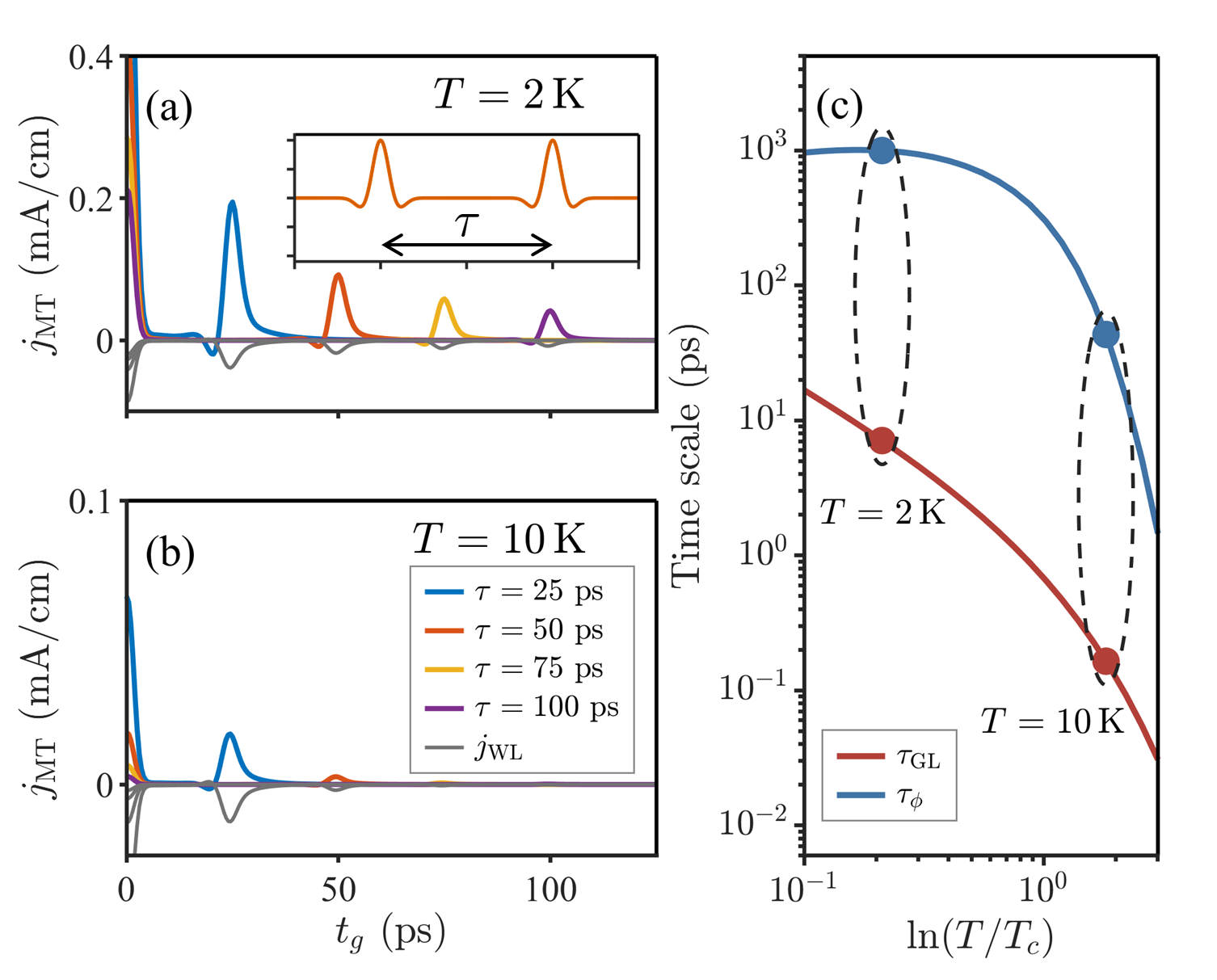}
\caption{The MT contribution $\bm{j}_{MT}$ in the aluminum film induced by two successive few cycle terahertz pulses. (a) The solid lines show $\bm{j}_{MT}$ as a function of the gating time $t_g$ for representative pulse delay time $\tau$ at $T=2$ K. The inset shows the pulse profile. (b) Same as (a), but for $T=10$ K. The gray lines in (a) and (b) are the corresponding weak localization contribution $j_{WL}$, which is much weaker than the MT contribution at $T=2$ K. (c) The Ginzburg-Landau time $\tau_{GL}$ and the phase coherence time $\tau_\phi$ as a function of temperature. The latter is taken from Ref.~\onlinecite{Gordon1986}.}
\label{fig:realistic}
\end{figure}

So far we have focused on the generic features of the photon echo. To make contact with experiments, we apply our theory to aluminum films using material parameters reported in Ref.~\cite{Gordon1986}. We set $D=10.8$ cm$^2$/s and $T_c=1.619$ K. The Ginzburg-Landau time $\tau_{GL}$ is calculated using the first expression of Eq.~\eqref{eq:tau_GL}. The phenomenological phase coherence time is calculated by using the empirical formula: 
\begin{align}
& 1/\tau_\phi [\textrm{ps}^{-1}] = \frac{2.111 \times 10^{-5}\, T[\textrm{K}]}{\ln(T/T_c)+0.011}
\nonumber\\
& \quad  + 3.272 \times 10^{-4}\, T[\textrm{K}] +1.968 \times 10^{-5}\, T^3 [\textrm{K}^3].
\end{align}
We use pulses of the form $\bm{E}(t) = \bm{E}_a \alpha(t) +\bm{E}_b \alpha(t-\tau)$, where the pulse profile $\alpha(t) = \exp(-\frac{t^2}{2 t^2_p}) \cos(2\pi f_0t)$. We set the central frequency $f_0 = 0.1$ THz to match the Planck time at $\sim O(1)$ Kelvin. The peak electric field strength $\bm{E}_a= 200$ V/cm and $\bm{E}_b=50$ V/cm. The pulse duration $t_p =2.121$ ps. 

We compute the current echo signal at two representative temperatures, one close to $T_c$ ($T=2$ K), as shown in Fig.~\ref{fig:realistic}(a), and the other far from $T_c$ ($T=10$ K), as shown in Fig.~\ref{fig:realistic}(b). The corresponding Ginzburg-Landau time $\tau_{GL}$ and the phase coherence time $\tau_\phi$ are marked in Fig.~\ref{fig:realistic}(c). For the pulse parameters considered here, $\bm{j}_{MT}$ reaches a magnitude on the order of $0.1$ mA/cm at $T=2$ K. It translates to a signal electric field $\bm{E}_{sig}\sim O(1)$ V/m, which is weak but detectable by electro-optical sampling with careful denoising and averaging~\cite{Szwaj2016}. Optimizing the pulse profile could further increase its magnitude although it is beyond the scope of the present work.

\section{Interference picture \label{sec:interference}}

\begin{figure}
\includegraphics[width = \columnwidth]{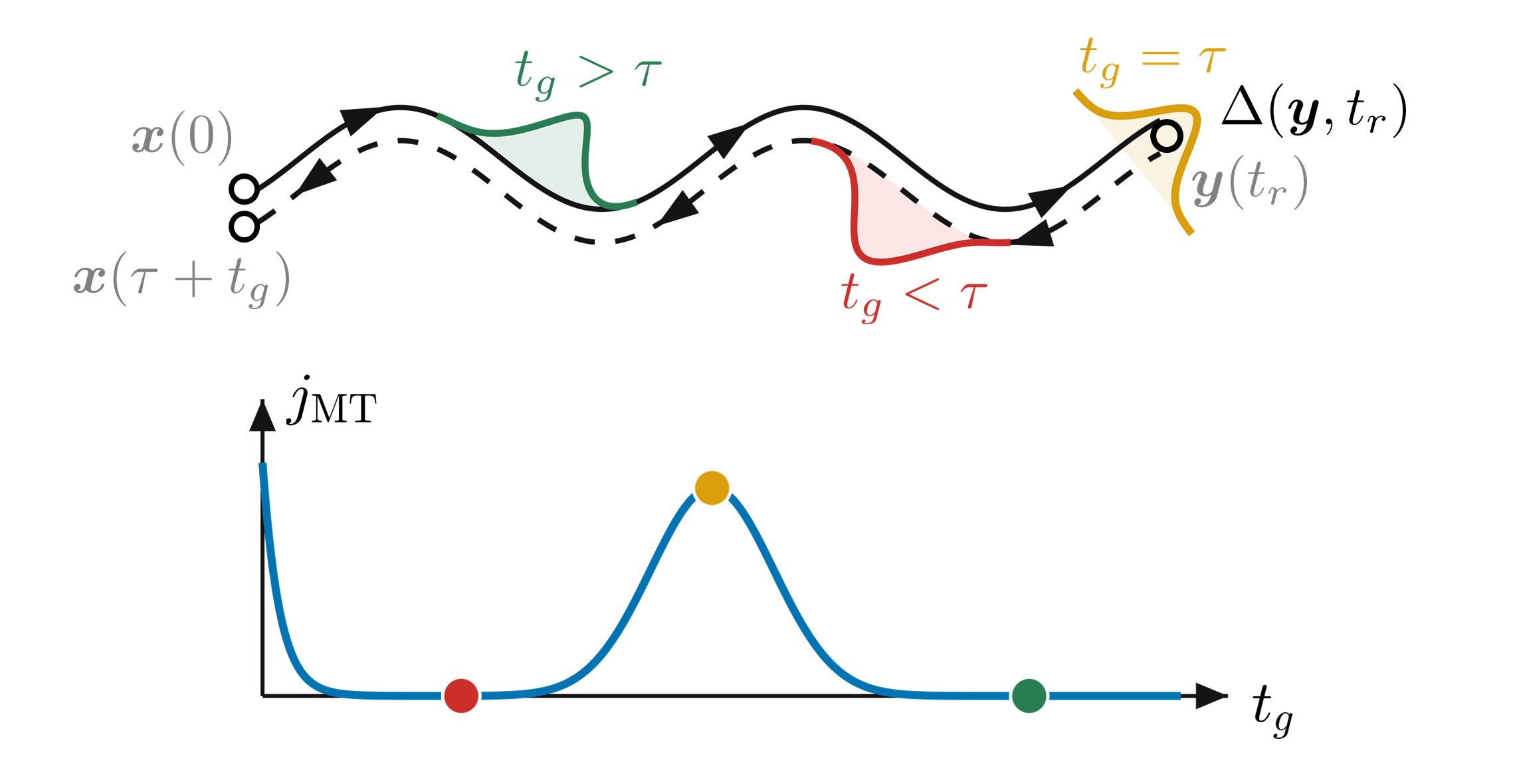}
\caption{The photon echo originates from the quantum interference process inherent to the MT contribution. In the two pulse set up considered here, the first pulse, anchored at time 0, initiates the interference process. An unparied electron starts at time zero, converts to a hole at the reflection time $t_r = (\tau+t_g)/2$, and returns at time $\tau+t_g$. The second pulse modulates the phase accumulated in the process. When $t_g \neq \tau$, the interference is suppressed, whereas, when $t_g \approx \tau$, the interference is restored. This suppression and restoration of interference gives rise to the photon echo.}
\label{fig:interference}
\end{figure}

In this section, we show that the photon echo signal in the MT contribution $\bm{j}_{MT}$ arises from the quantum interference between the time-reversed electron and hole trajectories in the MT process. 

To set the stage, we provide a heuristic derivation of Eq.~\eqref{eq:K_MT}. In the MT process, an unpaired electron propagates from the position $\bm{x}$ at time $t'$ to the position $\bm{y}$ at time $t_r$, gets converted to a hole by the superconducting order parameter $\Delta(\bm{y},t_r)$, and propagates back to the position $\bm{x}$ at time $t$. Its quantum mechanical amplitude reads:
\begin{align}
\mathcal{A}_{MT} \sim \int d^2 \bm{y} \int dt_r U_h(\bm{x},t;\bm{y},t_r) \Delta(\bm{y},t_r)
\nonumber\\
\times  U_e (\bm{y},t_r;\bm{x},t'),
\end{align}
where $U_e$ and $U_h$ are the electron and hole propagators, respectively. $t_r$ is the time of the reflection.

The electron propagator admits the following path integral representation:
\begin{align}
U_e (\bm{y},t_r;\bm{x},t') = \sum_{\bm{r}_e} e^{i\int ds [\frac{m\dot{\bm{r}}^2_e}{2} - V(\bm{r}_e) + e_0\bm{A}(s)\cdot\dot{\bm{r}}_e]}.
\end{align}
The electron trajectory $\bm{r}_e(s)$ is subject to the boundary conditions $\bm{r}_e(t') = \bm{x}$ and $\bm{r}_e(t_r) = \bm{y}$. $m$ is the effective mass of the electron. $V$ is the disorder potential. Meanwhile, the hole propagator is given by:
\begin{align}
U_h (\bm{x},t; \bm{y},t_r) = \sum_{\bm{r}_h} e^{-i\int ds [\frac{m\dot{\bm{r}}^2_h}{2} - V(\bm{r}_h) + e_0\bm{A}(s)\cdot\dot{\bm{r}}_h]}.
\end{align} 
The hole trajectory $\bm{r}_h$ is subject to the boundary conditions $\bm{r}_h(t_r) = \bm{y}$ and $\bm{r}_h(t) = \bm{x}$. The path integral representation for $U_h$ is deduced from that of $U_e$ by using the relation $U_h = U_e^\ast$. Alternatively, it can be obtained from the fact that the hole and the electron Hamiltonians are related by $H_h = -H^\ast_e$ in the Nambu spinor representation. 

We estimate the amplitude $\mathcal{A}_{MT}$ in the quasiclassical limit by invoking the stationary phase approximation. As the electron and hole path integrals have opposite signs in their respective phase factors, the amplitude $\mathcal{A}_{MT}$ is dominated by:
\begin{align}
\bm{r}_h (s) = \bm{r}_e(t+t'-s),
\end{align}
in the absence of the vector potential. In other words, the electron and the hole trajectories are time reversal partners. These trajectories satisfy the stationary condition because, when $\bm{A} = 0$, their total phase factor is exactly zero for \emph{any} disorder potential $V$. The contributions from other trajectories are suppressed by disorder averaging. An immediate consequence is that $t_r = (t+t')/2$, namely the reflection happens exactly halfway (Fig.~\ref{fig:interference}).

We anticipate that the same trajectory class dominates the MT amplitude even in the presence of a weak vector potential. In this case,
\begin{align}
\mathcal{A}_{MT}  \sim \int d^2\bm{y}\, \Delta(\bm{y},\frac{t+t'}{2}) \sum_{\bm{r}_e} e^{i\phi[\bm{r}_e]}
\end{align}
where the phase factor
\begin{align}
\phi[\bm{r}_e] = ie_0\int^{\frac{t+t'}{2}}_{t'} [\bm{A}(s) + \bm{A}(t+t'-s)]\cdot \dot{\bm{r}}_e ds.
\end{align}
The summation is over the electron trajectory; the hole trajectory is obtained from the electron trajectory by time reversal. If $\bm{A}(s)$ varies slowly compared with the typical time scale between two successive scattering events, $\tau_{el}$, the electron trajectory is well approximated by a continuous-time random walk from $\bm{x}$ at time $t'$ to $\bm{y}$ at time $(t+t')/2$. Thus, the path integral can be replaced by a statistical average over random walks:
\begin{align}
\mathcal{A}_{MT} &\sim \int d^2 \bm{y}\,  \Delta(\bm{y},\frac{t+t'}{2}) \langle e^{i\phi[\bm{r}_e]}\rangle_{D}
\nonumber\\
& = \int d^2 \bm{y}\,  \mathcal{C}^{\frac{t+t'}{2}}_{t-t',0}(x-y)\Delta(y,\frac{t+t'}{2}).
\end{align}
$\langle\cdots\rangle_{D}$ stands for the statistical average over the random walks with a diffusion constant $D$. In the second line, we have recognized that the statistical average yields the Cooperon~\cite{Altshuler1982}.

The MT correction to the conductivity is proportional to the probability of the MT process:
\begin{align}
& \langle |\mathcal{A}_{MT}|^2\rangle \sim \int d^2\bm{y}\, d^2\bm{z} \, \mathcal{C}^{\frac{t+t'}{2}}_{t-t',0}(\bm{x}-\bm{y})
\nonumber\\
&\quad \times \langle \Delta(\bm{y},\frac{t+t'}{2})\Delta^\ast(\bm{z},\frac{t+t'}{2})\rangle \overline{\mathcal{C}}^{\frac{t+t'}{2}}_{0,t-t'}(\bm{z}-\bm{x}).
\end{align}
Here, we have averaged over the fluctuations of the superconducting order parameter. The above is identical to Eq.~\eqref{eq:K_MT} except for a constant prefactor. 

The heuristic derivation provides a transparent interpretation of the effective vector potential $\bm{W}$ that appears in the coherence factor for the MT contribution (Eq.~\eqref{eq:W_def}). The physical vector potential $\bm{A}$ influences the MT amplitude $\mathcal{A}_{MT}$ through two distinct mechanisms. First, the vector potential $\bm{A}(s)$ drives the superconducting order parameter $\Delta(\bm{y},\frac{t+t'}{2})$ through the Ginzburg-Landau equation when $s<(t+t')/2$. Second, the vector potential modulates the interference between the electron and hole trajectories through the combination $\bm{A}(s) + \bm{A}(t+t'-s)$ when $s>(t+t')/2$. The different forms of the effective vector potential $\bm{W}(s)$ for $s<(t+t')/2$ and $s>(t+t')/2$ reflect these two mechanisms. 

We are now ready to clarify the origin of the photon echo based on the interference picture. In the two pulse protocol, we may regard the first pulse $\bm{E}_a$, anchored at time $0$, as initiating the MT process. The second pulse $\bm{E}_b$, arrives at time $\tau$, modulates the phase coherence between the electron and hole trajectories. At the detection time $\tau+t_g$, the hole propagates back to the initial position. Note that the reflection occurs at the time $t_r = (\tau+t_g)/2$. The phase due to the second pulse is given by:
\begin{align}
\phi \approx \left\{ \begin{array}{cc}
e_0 \sqrt{2\pi} t_p \bm{E}_b \cdot  \bm{x}_e (\tau)   & (t_g>\tau) \\
-e_0 \sqrt{2\pi} t_p \bm{E}_b \cdot   \bm{x}_e (t_g)  & (t_g<\tau) \\
0 & (t_g=\tau)
\end{array}\right. .
\end{align}
$\bm{x}_e$ is the position of the electron. $t_p$ is the pulse duration. Here, we have approximated the pulse as a Dirac delta function. When $t_g>\tau$ ($t_g<\tau$), the pulse acts on the electron (hole) trajectory. For both cases, $\phi$ is random because $\bm{x}_e$ is drawn from a random walk process. Thus, the phase coherence between the electron and hole trajectories is suppressed. However, when $t_g = \tau$, $\phi = 0$. The phase coherence is restored. This suppression and reinstatement of the phase coherence produces the photon echo when we scan $t_g$ past $\tau$ (Fig.~\ref{fig:interference}).

The interference picture naturally explains the decay of the echo signal. The phase information of the electron is maintained up to the phase coherence time $\tau_\phi$. Therefore, the quantum interference between the electron and hole trajectories is suppressed beyond that time scale. As a result, the photon echo disappears as the pulse delay time $\tau\gg\tau_\phi$.

\section{Discussion \label{sec:discussion}}

In summary, we have shown that the photon echo can qualitatively distinguish between the two main fluctuation conductivity channels in a fluctuating  s-wave superconductor. The MT process produces a photon echo when driven by two successive terahertz pulses, whereas the AL process does not. The echo is a direct manifestation of the quantum interference between the time-reversed electron and hole trajectories embedded in the MT process. Furthermore, the decay of the echo is a sensitive measure of electron dephasing in the time domain. 

Our model assumes a two-dimensional geometry. Extending the analysis beyond the two dimensional limit is straightforward. For instance, when the film thickness $d$ is larger than the superconducting correlation length $\xi_{GL}$ or phase-coherence length $l_\phi$, we need to replace the stochastic Ginzburg-Landau equations Eq.~\eqref{eq:TDGL} and the Cooperon equation Eq.~\eqref{eq:cooperon_equation} by their three-dimensional counterparts, supplemented with appropriate boundary conditions at the top and bottom surfaces. Furthermore, it will be necessary to couple them to the Maxwell equations because the vector potential inside the film varies in the thickness direction. Nonetheless, the qualitative features of the photon echo are robust as the quantum interference picture in Sec.~\ref{sec:interference} is valid in higher dimensions. 

Our analysis omits a weak localization contribution to the current response in Eq.~\eqref{eq:j_total}. This contribution is not specific to fluctuating s-wave superconductors but occurs in all disordered metals. A previous work has shown that the weak localization process can also produce a photon echo similar to that of the MT process through an analogous mechanism~\cite{Li2024}. Crucially, for the case considered here (orthogonal symmetry class, or negligible spin-orbit coupling), the echo from the weak localization contribution is \emph{negative}. In other words, it has an opposite sign to the echo from the MT contribution. Meanwhile, their temperature dependence is quite different. Therefore, even though both contributions produce an echo, it is possible to tell them apart when one contribution dominates over the other in appropriate temperature windows. For the aluminum example in Sec.~\ref{sec:echo}, our calculation shows that the weak localization echo is significantly weaker than the MT echo at temperatures close to $T_c$  (Figs.~\ref{fig:realistic}(a)).

Our work raises a few interesting open questions. First, we have treated dephasing phenomenologically by introducing an effective phase-coherence time, $\tau_\phi$, in the Cooperon equations Eq.~\eqref{eq:cooperon_equation}. In reality, dephasing due to inelastic scattering is frequency-dependent and therefore cannot, in general, be captured by a single constant~\cite{Altshuler1982,Marquardt2007,vonDelft2007}. Because our results show that the decay of the photon echo traces dephasing in the time domain, a more careful treatment of the inelastic scattering  will be necessary to obtain a quantitatively accurate echo decay profile. It may be possible to separate different dephasing mechanisms through their distinct decay profiles.

Second, our analysis is limited to the Gaussian fluctuations of the superconducting order parameter. It must be modified near the superconducting phase transition, where the non-Gaussian fluctuations become significant~\cite{Larkin2005}. This issue becomes particularly important in two dimensions where the phase fluctuations become important~\cite{Crane2007,Mondal2011}. We may partially account for the non-Gaussian fluctuations by replacing the linearized stochastic Ginzburg-Landau equations~\eqref{eq:TDGL} by the full, nonlinear version while retaining the expression for the current Eq.~\eqref{eq:j_total}~\cite{Larkin2005}. The MT contribution is corrected quantitatively through the order parameter correlation function $\langle \Delta(y,\frac{t+t'}{2}) \Delta^\ast (z,\frac{t+t'}{2})\rangle$ in Eq.~\eqref{eq:K_MT}. 

Finally, we note that the MT contribution to fluctuation conductivity is generally suppressed in superconductors with sign-changing unconventional pairing symmetries~\cite{Yip1990,Larkin2005}. We therefore expect the corresponding MT-induced photon echo to be suppressed as well. Nevertheless, it would be interesting to investigate quantum-interference processes in unconventional superconductors and to explore whether they can support alternative echo mechanisms. Beyond superconductivity, promising directions include systems with fluctuating charge-density-wave order and magnetic fluctuations.

\begin{acknowledgments}
This work is supported by the National Key R\&D development program (Grants No. 2024YFA1408700, 2022YFA1403800), the Natural Science Foundation of China (Grants No. 12547112, 12188101, 12250008), the CAS Project for Young Scientists in Basic Research (Grant No. YSBR-059), and the Postdoctoral Fellowship Program of the China Postdoctoral Science Foundation (Grant No. GZC20252262).
\end{acknowledgments}

\bibliography{fluc_sc_ref}

\end{document}